\documentclass[reqno,centertags, 12pt]{amsart}
\usepackage{amsmath,amsthm,amscd,amssymb}
\usepackage{latexsym}
\newcommand{\bbC}{{\mathbb{C}}}
\newcommand{\bbD}{{\mathbb{D}}}

\newcommand{\bbR}{{\mathbb{R}}}

\newcommand{\bbZ}{{\mathbb{Z}}}

\newcommand{\calC}{{\mathcal{C}}}

\newcommand{\calH}{{\mathcal H}}

\newcommand{\calL}{{\mathcal L}}
\newcommand{\calM}{{\mathcal M}}

\newcommand{\bdone}{{\boldsymbol{1}}}
\newcommand{\bddot}{{\boldsymbol{\cdot}}}

\newcommand{\lb}{\label}
\newcommand{\f}{\frac}

\newcommand{\ol}{\overline}
\newcommand{\ti}{\tilde  }

\newcommand{\loc}{\text{\rm{loc}}}

\newcommand{\ran}{\text{\rm{Ran}}}

\newcommand{\ac}{\text{\rm{ac}}}

\newcommand{\supp}{\text{\rm{supp}}}

\newcommand{\bi}{\bibitem}

\newcommand{\beq}{\begin{equation}}
\newcommand{\eeq}{\end{equation}}
\newcommand{\ba}{\begin{align}}
\newcommand{\ea}{\end{align}}
\newcommand{\veps}{\varepsilon}

\newcommand{\fre}{\frak{e}}
\newcommand{\pj}{P_{\Sigma_{\text{\rm{ac}}}^{(2)}}}

\newcommand{\abs}[1]{\lvert#1\rvert}
\newcommand{\jap}[1]{\langle #1 \rangle}
\newcommand{\norm}[1]{\lVert#1\rVert}

\newcounter{smalllist}

\DeclareMathOperator{\Real}{Re}
\DeclareMathOperator{\Ima}{Im}

\DeclareMathOperator*{\slim}{s-lim}

\allowdisplaybreaks
\numberwithin{equation}{section}

\newtheorem{theorem}{Theorem}[section]
\newtheorem*{mt}{Almost-Theorem 1.6}

\newtheorem{corollary}[theorem]{Corollary}
\theoremstyle{definition}

\theoremstyle{remark}
\newtheorem*{remark}{Remark}

\begin{document}
\title[Spectral and Dynamical Reflection]{Equality of the Spectral and Dynamical Definitions of Reflection}
\author[J.~Breuer, E.~Ryckman, and B.~Simon]{Jonathan Breuer, Eric Ryckman,
and Barry Simon$^*$}

\thanks{Mathematics 253-37, California Institute of Technology, Pasadena, CA 91125, USA.
E-mail: jbreuer@caltech.edu; eryckman@caltech.edu;
bsimon@caltech.edu.}

\thanks{$^*$ Supported in part by NSF grant DMS-0652919}

\date{May 12, 2009}


\begin{abstract}
For full-line Jacobi matrices, Schr\"odinger operators, and CMV matrices, we show that being reflectionless, in the sense
of the well-known property of $m$-functions, is equivalent to a lack of reflection in the dynamics in the sense that any state
that goes entirely to $x=-\infty$ as $t\to -\infty $ goes entirely to $x=\infty$ as $t\to\infty$. This allows us to settle
a conjecture of Deift and Simon from 1983 regarding ergodic Jacobi matrices.
\end{abstract}

\maketitle

\section{Introduction} \lb{s1}

In this paper, we discuss dynamics and spectral theory of whole-line Jacobi matrices, Schr\"odinger
operators, and CMV matrices.  In this introduction we focus on Jacobi matrices, that is, doubly
infinite matrices,
\begin{equation} \lb{1.1}
J = \begin{pmatrix}
\ddots & \ddots  & \ddots \\
{} & a_{-2} & b_{-1} & a_{-1} \\
{} & {} & a_{-1} & b_0 & a_0 \\
{} & {} & {} & a_0 & b_1 & a_1 \\
{} & {} & {} & {} & \ddots & \ddots & \ddots
\end{pmatrix}
\end{equation}
acting as operators on $\ell^2(\bbZ)$. We suppose throughout that the Jacobi parameters,
$\{a_n,b_n\}_{n=-\infty}^\infty$, are bounded.

We will sometimes need half-line Jacobi matrices given by
\begin{equation} \lb{1.1a}
\begin{pmatrix}
b_1 & a_1 & 0 & \dots \\
a_1 & b_2 & a_2 & \dots \\
0 & a_2 & b_3 & \dots \\
\vdots & \vdots & \vdots & \ddots
\end{pmatrix}
\end{equation}
We call $\{a_n,b_n\}_{n=1}^\infty$ the Jacobi parameters for a half-line matrix and
$\{a_n,b_n\}_{n=-\infty}^\infty$ the Jacobi parameters for a whole-line matrix.

We will call $J$ {\it measure theoretically reflectionless\/} on a Borel set $\fre\subset\bbR$
if and only if for all $n$, the diagonal Green's function,
\begin{equation} \lb{1.1c}
G_{nn} (\lambda+i0) =\lim_{\veps\downarrow 0} \jap{\delta_n,
(J-\lambda-i\veps)^{-1}\delta_n}
\end{equation}
is pure imaginary, that is,
\begin{equation} \lb{1.1d}
\Real G_{nn}(\lambda+i0)=0
\end{equation}
for Lebesgue a.e.\ $\lambda\in\fre$.
Among the vast literature, we mention
\cite{Cr89,DJ87,DS83,GKT96,GMZ08,GY06,GZ08,Jo82,Ko84,Ko87a,KK88,MPV08,NVY08,PY03,PR08,PSZ,RemSc,Rem,Si07a,SY95,SY95a,SY96,SY97}.
The name ``reflectionless'' is usually used without ``measure
theoretically'' but we add this for reasons that will be clear
shortly.

The notion first became commonly used in connection with solitons and has recently become especially
important because of Remling's discovery \cite{Rem} that right limits of half-line Jacobi matrices are
measure theoretically reflectionless on $\Sigma_\ac$, the essential support of the a.c.\ component of
the half-line Jacobi matrix. The name comes from the fact that in the short-range case (i.e., $\abs{a_n-1}
+ \abs{b_n}\to 0$ sufficiently rapidly as $\abs{n}\to\infty$), the condition is equivalent to the
time-independent reflection coefficient being zero on $\fre$.

There is a second notion of reflectionless operator
 depending on ideas of Davies--Simon
\cite{Da-Si}. For each $n\in\bbZ$, let $\chi_n^+$ be the
characteristic function of $[n,\infty)$ and $\chi_n^-$ of
$(-\infty,n]$. We define
\begin{equation} \lb{1.1b}
\calH_\ell^+ = \bigl\{\varphi\in\calH_\ac\bigm| \text{for all } n,\, \lim_{t\to -\infty}\,
\norm{\chi_n^+ e^{-itJ} \varphi}=0\bigr\}
\end{equation}
that is, states that, as $t\to-\infty$, are concentrated on the left. $\calH_\ell^-$ is the same with
$\lim_{t\to +\infty}$, and $\calH_r^\pm$ are defined using $\chi_n^-$. Here $\calH_\ac$ is the a.c.\ subspace
for $J$. We let $P_\ac$ be the projection onto $\calH_\ac$, and let $P_{\ell,r}^\pm$ be the orthogonal
projection onto $\calH_{\ell,r}^\pm$, that is,
\begin{equation}\label{projection definition}
P_\ell^\pm =\slim_{t\to \mp \infty}\, e^{itJ} \chi_0^- e^{-itJ} P_\ac (J) \qquad
P_r^\pm =\slim_{t\to \mp \infty}\, e^{itJ} \chi_0^- e^{-itJ} P_\ac (J)
\end{equation}

Davies--Simon prove (they treat the analog for Schr\"odinger operators, but the argument
is identical):

\begin{theorem}[\cite{Da-Si}] \lb{T1.1} We have {\rm{(}}$\oplus =$ orthogonal direct sum{\rm{)}}
\begin{align}
\calH_\ac &= \calH_\ell^+ \oplus \calH_r^+ \lb{1.2} \\
&= \calH_\ell^- \oplus \calH_r^- \lb{1.3}
\end{align}
That is, any a.c.\ state is a sum of a state that moves entirely to the left as $t\to -\infty$ and
one that moves to the right.
\end{theorem}

We call $J$ {\it dynamically reflectionless\/} on a Borel set $\fre$ if and only if
\begin{equation} \lb{1.7a}
P_\fre P_\ac = P_\fre
\end{equation}
(here $P_\fre$ is the spectral projection for $J$) and
\begin{equation} \lb{1.4}
P_\fre [\calH_\ell^+] = P_\fre [\calH_r^-]
\end{equation}

Before stating our main theorem, we want to define a third notion of
reflectionless operator  for reasons that will
become clear momentarily. For any $n\in\bbZ$, let $J_n^+$ be the
Jacobi matrix obtained from dropping the row and column with $b_n$
and keeping the lower right piece, that is, $J_n^+$ is the one-sided
Jacobi matrix with Jacobi parameters
\begin{equation} \lb{1.5}
b_\ell^{(n),+} = b_{n+\ell} \qquad a_\ell^{(n),+} = a_{n+\ell}
\end{equation}
$J_n^-$ has parameters
\begin{equation} \lb{1.6}
b_\ell^{(n),-} = b_{n+1-\ell} \qquad a_\ell^{(n),-} = a_{n-\ell}
\end{equation}

Thus, if $a_n$ is replaced by $0$, the whole-line Jacobi matrix $J$ breaks into a direct sum of $J_n^+$ and a matrix
unitarily equivalent to $J_n^-$ after reordering the indices in inverse order.

For any half-line Jacobi matrix, $J$, we define its $m$-function by
\begin{equation} \lb{1.7}
m(z,J) = \jap{\delta_1, (J-z)^{-1}\delta_1}
\end{equation}
and for a whole-line Jacobi matrix,
\begin{equation} \lb{1.8}
m_n^\pm (z,J) = m(z,J_n^\pm)
\end{equation}
These are related to the Green's function \eqref{1.1c} by
\begin{equation} \lb{1.9}
G_{nn}(z) = -\f{1}{a_n^2 m_n^+(z) - m_n^-(z)^{-1}}
\end{equation}
We call a whole-line Jacobi matrix {\it spectrally reflectionless\/}
on a Borel set $\fre$ if for a.e.\ $\lambda\in\fre$ and all $n$,
\begin{equation} \lb{1.10}
a_n^2 m_n^+(\lambda+i0) \, \ol{m_n^- (\lambda+i0)} =1
\end{equation}

By \eqref{1.9}, \eqref{1.10} implies $\Real G_{nn} =0$, so
\[
\text{\eqref{1.10} for $\lambda$ and $n \Rightarrow$ \eqref{1.1d}
for $\lambda$ and $n$}
\]
and so
\begin{equation}  \lb{1.11} \begin{split}
& \text{$J$ is spectrally reflectionless on $\fre$} \Rightarrow \\ &
\text{$J$ is measure theoretically reflectionless on $\fre$}
\end{split}
\end{equation}

Moreover, as we will see below,
\begin{equation} \lb{1.11a}
\text{\eqref{1.10} for $\lambda$ and one $n \Rightarrow$
\eqref{1.10} for $\lambda$ and all $n$}
\end{equation}
This set of ideas is rounded out by the following theorem:

\begin{theorem}[Gesztesy--Krishna--Teschl \cite{GKT96}; Sodin--Yuditskii \cite{SY97}] \lb{T1.2} If \eqref{1.1d}
holds for a.e.\ $\lambda\in\fre$ and three consecutive values of
$n$, then \eqref{1.10} holds for a.e.\ $\lambda\in\fre$ and all $n$.
\end{theorem}

In particular, in \eqref{1.11}, $\Rightarrow$ can be replaced by $\Leftrightarrow$. However, this is not
true for CMV matrices \cite{BRZ}.

Here is our main result:

\begin{theorem} \lb{T1.3} For any whole-line Jacobi matrix $J$ and Borel set $\fre$ of positive Lebesgue
measure, $J$ is spectrally reflectionless on $\fre$ if and only if it is dynamically reflectionless
on $\fre$.
\end{theorem}

This verifies a 25-year old conjecture of Deift--Simon \cite{DS83}, namely

\begin{corollary} \lb{C1.4} The a.c.\ spectrum for two-sided ergodic Jacobi matrices is dynamically
reflectionless.
\end{corollary}

\begin{proof} By Kotani theory \cite{Ko84,S168}, such operators are spectrally reflectionless on
the a.c.\ spectrum.
\end{proof}

This is a special case of a more general result that we will prove
concerning reflection probability. Let $\Sigma_\ac^{(2)}$ be the set
of $\lambda\in\bbR$ where $J$ has multiplicity $2$,
so automatically a.c.\ spectrum (see
\cite{Gil98,IKac62,IKac,S293}). $P_{\ell,r}^\pm$ commute with $J$,
so they take $\ran (\pj(J))$ to itself. $J$ restricted to $\ran
(P_{\ell,r}^\pm \pj (J))$ is of multiplicity $1$. Thus,
\begin{equation} \lb{1.12}
R=P_\ell^+ P_\ell^- P_\ell^+ \restriction \ran (P_\ell^+ \pj (J))
\end{equation}
is a scalar function of $J$, and so there is a function $R(E)$ on $\Sigma_\ac^{(2)}$ so that
\begin{equation} \lb{1.13}
R=R(J)\restriction \ran(P_\ell^+ \pj (J))
\end{equation}
As defined by Davies--Simon \cite{Da-Si}, $R(\lambda)$ is the {\it dynamic reflection probability},
the probability that a state of energy $\lambda$ that comes in from the left at very negative
times goes out on the left. There is a time-reversal symmetry,
namely that one gets the same function, $R$, with $P_\ell^- P_\ell^+
P_\ell^- \restriction \ran (P_\ell^-)$. Similarly, there is a
left-right symmetry, so one gets the same function with $P_r^+ P_r^-
P_r^+ \restriction \ran (P_r^+)$.

Define the {\it spectral reflection probability\/} by (see Theorem~\ref{T2.5} below for why this is
a good definition)
\begin{equation} \lb{1.14}
\biggl| \f{a_0^2 m_0^+ (\lambda+i0)\, \ol{m_0^- (\lambda-i0)} -1}
{a_0^2 m_0^+ (\lambda+i0) m_0^- (\lambda-i0)-1} \biggr|^2
\end{equation}
We will prove

\begin{theorem}\lb{T1.5} $R(\lambda)$ is given by \eqref{1.14} on $\Sigma_\ac^{(2)}(J)$.
\end{theorem}

Theorem~\ref{T1.5} implies Theorem~\ref{T1.3} since
\begin{equation} \lb{1.15}
R(J) \restriction \fre =0 \Leftrightarrow P_\ell^+ P_\fre  = P_r^- P_\fre
\end{equation}
and
\begin{equation} \lb{1.16}
\text{\eqref{1.14} $=0 \Leftrightarrow$ \eqref{1.10} holds}
\end{equation}

The various formulae involving $m_n^\pm$ are complicated, in part because the simple formulae
are given by Weyl solutions. It pays to rewrite them here since the rewriting is critical to
our proof.

We are interested in solutions of
\begin{equation} \lb{1.17}
a_{n-1} u_{n-1} + b_n u_n + a_n u_{n+1} = zu_n
\end{equation}
For any $z\in\bbC_+ =\{z\mid\Ima z>0\}$, there are solutions $u_n^\pm (z)$ which are $\ell^2$ at
$\pm\infty$, unique up to a constant. We will normalize by
\begin{equation} \lb{1.17a}
u_0^\pm =1
\end{equation}
By general principles (see, e.g., \cite[Chap.~2]{Teschl},
 though our notation
is slightly different from his), for Lebesgue a.e.\ $\lambda$,
$u_n^\pm (\lambda+i\veps)$ has a limit as $\veps \downarrow 0$,
which we denote by $u_n^\pm (\lambda+i0)$ which solves \eqref{1.17}
at $\lambda$.

$m^\pm$ can be expressed in terms of $u^\pm$ by (\cite{Teschl})
\begin{align}
m_n^+ (\lambda+i0) &= - \f{u_{n+1}^+ (\lambda+i0)}{a_n u_n^+ (\lambda+i0)} \lb{1.18} \\
m_n^- (\lambda+i0) &= - \f{u_n^- (\lambda+i0)}{a_n u_{n+1}^-
(\lambda+i0)} \lb{1.19}
\end{align}
The Green's function, \eqref{1.1c}, which is symmetric, is given for $n\leq m$ by
\begin{equation} \lb{1.20}
G_{nm} (\lambda+i0) = \f{u_n^- (\lambda+i0) u_m^+
(\lambda+i0)}{W(\lambda+i0)}
\end{equation}
where
\begin{equation} \lb{1.21}
W(z) = a_n [u_{n+1}^+(z) u_n^-(z) - u_{n+1}^-(z) u_n^+(z)]
\end{equation}
is $n$-independent.

From these formulae, \eqref{1.9} is immediate. Moreover, with the
normalization $u_{n=0}^\pm =1$, we see that \eqref{1.10} is
equivalent to $u_{n=1}^+ (\lambda+i0) = \ol{u_{n=1}^- (\lambda+i0)}$
which, by uniqueness of solutions, implies
\begin{equation} \lb{1.22}
u_n^+ (\lambda+i0) = \ol{u_n^- (\lambda+i0)}
\end{equation}
for all $n$. This explains why \eqref{1.11a} holds. It shows that
\begin{equation} \lb{1.23}
\text{$J$ is spectrally reflectionless for $\lambda\in\fre
\Leftrightarrow$ \eqref{1.22} for $\lambda\in\fre$}
\end{equation}

The key to our proof of Theorem~\ref{T1.3} (and also Theorem~\ref{T1.5}) will be

\begin{mt} $\ran(P_\ell^+ \pj)$ is spanned by $\{u_n^+ (\lambda+i0)\mid \lambda\in \Sigma_\ac^{(2)}\}$ and
$\ran(P_r^+ \pj )$ by $\{u_n^- (\lambda+i0) \mid
\lambda\in\Sigma_\ac^{(2)}\}$.
\end{mt}

We call this an almost-theorem because we are, for now, vague about
what we mean by ``span.'' The $u_n^\pm$ are only continuum
eigenfunctions, so by span we will mean suitable integrals.

We can now understand why the almost-theorem will imply
Theorem~\ref{T1.3}. By time-reversal invariance,
\begin{equation} \lb{1.24}
P_r^- = \ol{P_r^+}
\end{equation}
Thus,
\begin{equation} \lb{1.25}
\text{$J$ is dynamically reflectionless for } \lambda\in\fre
\Leftrightarrow \ol{P_r^+}\, \pj = P_\ell^+ \pj
\end{equation}
and the almost-theorem says the right side is the same as
\eqref{1.22}.

For short-range perturbations of the free Jacobi matrix ($b_n\equiv
0$, $a_n\equiv 1$), the almost-theorem follows from suitable
stationary phase/integration by parts ideas as noted in
Davies--Simon \cite{Da-Si}. Such methods cannot work for general
Jacobi matrices where $\Sigma_\ac^{(2)}$ might be a positive measure
Cantor set. What we will see is by replacing the limit
\begin{equation} \lb{1.26}
P_\ell^+ =\slim_{t\to -\infty}\, e^{itJ} \chi_0^- e^{-itJ} P_\ac (J)
\end{equation}
that Davies--Simon \cite{Da-Si} use by an abelian limit, a simple
calculation will yield the almost-theorem.

Section~\ref{s2} proves all the above results for Jacobi matrices. Section~\ref{s3} discusses (continuum)
Schr\"odinger operators and Section~\ref{s4} CMV matrices.

\section{The Jacobi Case} \lb{s2}

In this section, we prove Almost-Theorem~1.6 and use it to prove
Theorem~\ref{T1.5}, and thereby Theorem~\ref{T1.3}. To make sense of
Almost-Theorem~1.6, we need to begin with an eigenfunction
expansion. While this expansion can be viewed as a rephrasing of
Section~2.5 of Teschl \cite{Teschl}, it is as easy to establish it
from first principles as to manipulate the results of \cite{Teschl}
to the form we need. Our use of Stone's formula is similar to that
of Gesztesy--Zinchenko \cite{GZ06}.

Fundamental to this is the matrix for $\lambda\in\bbR$,
\begin{equation} \lb{2.1}
S(\lambda)_{nm}=\lim_{\veps\downarrow 0}\, (2\pi i)^{-1} [(J-\lambda-
i\veps)^{-1} - (J-\lambda+i\veps)^{-1}]_{nm}
\end{equation}
defined for a.e.\ $\lambda\in\bbR$ and all $n,m$. We use $S$ for
``Stone'' or ``spectral'' since Stone's formula (Thm.~VII.13 of
\cite{RS1}) and the spectral theorem imply that for any
$\varphi,\psi$ of finite support on $\bbZ$ and any Borel set,
$\fre$,
\begin{equation} \lb{2.2x}
\jap{\varphi, P_\fre P_\ac \psi} = \int_{\lambda\in\fre} \biggl(\,
\sum_{n,m} \bar\varphi_n \psi_m S(\lambda)_{nm} \biggr)\, d\lambda
\end{equation}

Define for $\lambda\in\Sigma_\ac^{(2)}$,
\begin{equation} \lb{2.2}
f_\pm (\lambda) = \pm \f{a_0 \Ima (u_1^\mp
(\lambda+i0))}{\pi\abs{W(\lambda+i0)}^2}
\end{equation}
where $u_n^\pm$ is normalized by \eqref{1.17a} and $W$ is given by \eqref{1.21}. This looks asymmetric
in $\pm$, but
\begin{align}
f_+(\lambda) &= -\f{a_{-1} \Ima(u_{-1}^- (\lambda+i0))}{\pi\abs{W(\lambda+i0)}^2} \lb{2.3} \\
&= \f{a_{-1}^2 \Ima (m_{-1}^- (\lambda+i0))}{\pi\abs{W(\lambda+i0)}^2} \lb{2.4} \\
\intertext{while} f_-(\lambda) &= \f{a_0^2 \Ima (m_0^+
(\lambda+i0))}{\pi\abs{W(\lambda+i0)}^2} \lb{2.5}
\end{align}
symmetric under reflection about $n=0$. This makes it clear that
\begin{equation} \lb{2.6}
f_\pm(\lambda) >0 \qquad \text{a.e. } \lambda\in \Sigma_\ac^{(2)}
\end{equation}

The key to our eigenfunction expansion is
\begin{equation} \lb{2.7}
S_{nm}(\lambda) = \ol{u_n^+ (\lambda+i0)}\, u_m^+ (\lambda+i0) f_+
(\lambda) + \ol{u_n^- (\lambda+i0)}\, u_m^- (\lambda+i0)
f_-(\lambda)
\end{equation}
for all $n,m$ and a.e.\ $\lambda\in\Sigma_\ac^{(2)}$.

\begin{theorem}\lb{T2.1} \eqref{2.7} holds for all $n,m$ and a.e.\ $\lambda\in\Sigma_\ac^{(2)}$.
\end{theorem}

\begin{proof} By general principles on limits of Stieltjes transforms, for a.e.\ $\lambda\in\Sigma_\ac^{(2)}$,
$\lim_{\veps\downarrow 0} u_n^\pm (\lambda+i\veps) = u_n^\pm
(\lambda+i0)$ exists. We will prove \eqref{2.7} for such $\lambda$.
It is easy to see that
$S_{nm}(\lambda)=S_{mn}(\lambda)$, so it suffices to consider the case $n\leq m$.

By the resolvent formula, for $\Ima z >0$,
\begin{align}
\pi S_{nm}(z) &\equiv (2i)^{-1} [(J-z)^{-1} - (J-\bar z)^{-1}]_{nm} \notag \\
&= (\Ima z) \sum_k (J-\bar z)_{nk}^{-1} (J-z)_{km}^{-1} \lb{2.9} \\
&= (\Ima z) \abs{W(z)}^{-2} (t_{nm}^{(1)} + t_{nm}^{(2)} + t_{nm}^{(3)}) \lb{2.10}
\end{align}
by \eqref{1.20}, where
\begin{align}
t_{nm}^{(1)} &=\biggl[\,\sum_{k\leq n} \, \abs{u_k^-(z)}^2\biggr] \ol{u_n^+(z)}\, u_m^+(z) \lb{2.11} \\
t_{nm}^{(2)} &=\biggl[\, \sum_{k\geq m+1} \, \abs{u_k^+(z)}^2\biggr] \ol{u_n^-(z)}\, u_m^-(z) \lb{2.12} \\
t_{nm}^{(3)} &=\biggl[\, \sum_{k=n+1}^m\,  \ol{u_k^+(z)}\, u_k^-(z)\biggr] \ol{u_n^-(z)}\,u_m^+(z) \lb{2.13aa}
\end{align}

Because of the $\Ima z$ in front of \eqref{2.10}, $\lim(\Ima
z)t_{nm}^{(3)}(\lambda+iy)=0$ since the limit exists (the sum is
finite).  Similarly, we can change
the summation limits of the $k$ sums in $t^{(1)},t^{(2)}$ to any
other finite value, since in the limit, finite sums multiplied by
$\Ima z$ go to zero. The result is
\begin{equation} \lb{2.13a}
S_{nm}(\lambda+i0) = q^{(1)}(\lambda)\, \ol{u_n^+ (\lambda+i0)}\,
u_m^+ (\lambda+i0) + q^{(2)}(\lambda)\, \ol{u_n^-(\lambda+i0)}\,
u_m^- (\lambda+i0)
\end{equation}
where
\begin{align}
\pi q^{(1)}(\lambda) &= \lim_{\veps\downarrow 0}\,
\abs{W(\lambda+i0)}^{-2} \veps \sum_{k\leq -1}\,
\abs{u_k^- (\lambda+i\veps)}^2 \lb{2.13.b} \\
\pi q^{(2)}(\lambda) &= \lim_{\veps\downarrow 0}\,
\abs{W(\lambda+i0)}^{-2} \veps \sum_{k\geq 1}\, \abs{u_k^+
(\lambda+i0)}^2 \lb{2.13.c}
\end{align}

By the resolvent formula for $J_0^+$ and the analog of \eqref{1.20} (with the normalization \eqref{1.17a}),
\begin{align}
\Ima m_0^+ (z) &= \Ima (J_0^+ -z)_{11}^{-1} \notag \\
&= (\Ima z) \sum_{k=1}^\infty (J_0^+ -\bar z)_{1k}^{-1} (J_0^+ -z)_{k1}^{-1} \notag \\
&= (\Ima z) a_0^2 \sum_{k=1}^\infty\, \abs{u_k^+(z)}^2 \lb{2.13d}
\end{align}
so
\begin{equation} \lb{2.13e}
q^{(2)}(\lambda) = f_-(\lambda)
\end{equation}
and similarly,
\begin{equation} \lb{2.13f}
q^{(1)}(\lambda) = f_+(\lambda)
\end{equation}

This proves \eqref{2.7}.
\end{proof}

From \eqref{2.7}, we immediately get an eigenfunction expansion.

\begin{theorem}\lb{T2.3} For any $\varphi\in\ell^2(\bbZ)$ of finite support, define
\begin{equation} \lb{2.13}
\widehat\varphi_\pm(\lambda) = \sum_n\, \ol{u_n^\pm(\lambda)}\,
\varphi_n
\end{equation}
as functions on $\Sigma_\ac^{(2)}$. Then
\begin{equation} \lb{2.14}
\int_{\Sigma_\ac^{(2)}} [\abs{\widehat\varphi_+(\lambda)}^2
f_+(\lambda) + \abs{\widehat\varphi_-(\lambda)}^2 f_-(\lambda)]\,
d\lambda = \norm{P_\ac \pj \varphi}^2
\end{equation}
So $\,\widehat{\,\,}_\pm$ extend to continuous maps of
$\ell^2(\bbZ)$ to $L^2 (\Sigma_\ac^{(2)}, f_\pm\, d\lambda)$.
Moreover, if $\,\widehat{\,\,} = (\,\widehat{\,\,}_+,
\,\widehat{\,\,}_-)$, then
\begin{equation} \lb{2.15}
\widehat{(J\varphi)}_\pm(\lambda) = \lambda \widehat\varphi_\pm
(\lambda)
\end{equation}

For each $n$,
\begin{equation} \lb{2.15a}
\int_{\Sigma_\ac^{(2)}} \abs{u_n^\pm(\lambda)}^2 f_\pm(\lambda)\,
d\lambda \leq 1
\end{equation}

In particular, for any
\[
g=(g_+, g_-)\in L^2 (\Sigma_\ac^{(2)}, f_+\, d\lambda) \oplus L^2
(\Sigma_\ac^{(2)}, f_-\, d\lambda) \equiv \calH_J
\]
and any $n$, we can define
\begin{equation} \lb{2.15b}
\check{g}_n = \int g_+(\lambda) u_n^+ (\lambda+i0) f_+(\lambda)\,
d\lambda + \int g_-(\lambda) u_n^-(\lambda+i0) f_-(\lambda)\,
d\lambda
\end{equation}
$\check{g}$ lies in $\ell^2(\bbZ)$, and for any $\varphi\in\ell^2$,
\begin{equation} \lb{2.15c}
\jap{\check{g},\varphi} = \jap{g,\widehat\varphi}
\end{equation}
and
\begin{equation} \lb{2.15d}
\widehat{\check{g}} =g
\end{equation}

We have $\check{g}\in\ran(P_\ac  \pj)$ and $\check{\,}$ is a bijection of this range and $\calH_J$.
\end{theorem}

\begin{proof} \eqref{2.14} is immediate from \eqref{2.2x} and \eqref{2.7}. \eqref{2.15} follows
from a summation by parts and
\begin{equation} \lb{2.16}
\sum_m J_{nm} u_m^\pm (\lambda+i0) = \lambda u_n^\pm (\lambda+i0)
\end{equation}
\eqref{2.15a} comes from putting $\delta_n$ into \eqref{2.14}.

By \eqref{2.15a}, the integrals in \eqref{2.15b} converge for all $g\in\calH_J$. For $\varphi$ of
finite support, \eqref{2.15c} is an interchange of integration and finite sum. In particular, if $\chi_N$
is the characteristic function of $\{j\in\bbZ\mid \abs{j}\leq N\}$ and $\varphi =\chi_N \check{g}$,
\eqref{2.15c} implies
\begin{align}
\sum_{\abs{j}\leq N}\, \abs{\check{g}_j}^2
&\leq \norm{g}\, \norm{\widehat\varphi} \notag \\
&\leq \norm{g}\, \norm{\varphi} \notag \\
&= \norm{g} \biggl(\, \sum_{\abs{j} \leq N}\, \abs{g_j}^2\biggr)^{1/2} \lb{2.17}
\end{align}
so for all $N$,
\begin{equation} \lb{2.18}
\norm{\chi_N \check{g}} \leq \norm{g}
\end{equation}
so $\check{g}\in\ell^2$ and
\begin{equation} \lb{2.19}
\norm{\check{g}} \leq \norm{g}
\end{equation}
Thus, \eqref{2.15c} extends to all $\varphi$ by continuity.

By \eqref{2.14} and \eqref{2.15}, $\widehat{\,\,}\,$ is a unitary spectral representation for $\ti J=J
\restriction\ran(P_\ac \pj)$ on $\ran(\,\widehat{\,\,}\,)$. Since $\ti J$ has uniform multiplicity $2$,
$\ran(\,\widehat{\,\,}\,)$ must be all $\calH_J$. It follows that $(\,\widehat{\,\,}\,)(\,\widehat{\,\,}\,)^*
=\bdone$ on $\calH_J$. Since $\check{\,\,}=(\,\widehat{\,\,}\,)^*$, this is \eqref{2.15d}.
\end{proof}

We can now prove a precise version of Almost-Theorem~1.6. Let
\begin{equation} \lb{2.20}
\calH_J^\pm = \{g\in\calH_J \mid g_{\mp}=0\}
\end{equation}
and let $P^\pm$ be the projection in $\ell^2 (\bbZ)$ onto the image of $\calH_J^\pm$ under $\check{\,\,}$.
Then

\begin{theorem} \lb{T2.4} We have
\begin{align}
P_r^+ \pj (J) &= P^- \lb{2.21} \\
P_\ell^+ \pj (J) &= P^+ \lb{2.22} \\
P_r^- \pj (J) &= \ol{P^-} \lb{2.23} \\
P_\ell^- \pj (J) &= \ol{P^+} \lb{2.24}
\end{align}
\end{theorem}

\begin{remark} Let $C$ be complex conjugation on $\ell^2$. By $\bar A$ we mean $C\!AC$.
\end{remark}

\begin{proof} We claim that it suffices to prove for $\varphi\in\ran (P^+)$ that
\begin{equation} \lb{2.25}
P_\ell^+ \varphi = \varphi
\end{equation}
for then, by reflection in $n=0$, we see that for $\psi\in\ran (P^-)$,
\begin{equation} \lb{2.26}
P_r^+ \psi = \psi
\end{equation}
and
\begin{equation} \lb{2.27}
(P_\ell^+ + P_r^+) \pj (J) = \pj (J) = P^+ + P^-
\end{equation}
implies \eqref{2.21}/\eqref{2.22}. Since $\ol{e^{-itJ}} = e^{itJ}$, \eqref{2.23}/\eqref{2.24} then follow.

Clearly, it suffices to prove \eqref{2.25} for a dense set of
$\varphi\in \ran(P^+)$; equivalently, for a dense set of $g\in L^2
(\Sigma_\ac^{(2)}, f_+\, d\lambda)$, where
\begin{equation} \lb{2.28}
\varphi_n = \int g(\lambda) u_n^+ (\lambda+i0) f_+(\lambda)\,
d\lambda
\end{equation}

By Egoroff's theorem, for a dense set of $g$, we can suppose $g\in
L^\infty$, and for each fixed $m,n$, $G_{nm}(\lambda + ik^{-1}) \to
G_{nm}(\lambda+i0)$ as $k\to\infty$, uniformly for
$\lambda\in\supp(g)$. We henceforth assume these properties for $g$.

By \eqref{projection definition} and an abelian theorem \cite[Sect.~XI.6, Lemma~5]{RS3},
\begin{align*}
P_\ell^+ \varphi
&= \chi_0^- \varphi - i\, \lim_{t\to -\infty} \int _t^0 e^{isJ} [J,\chi_0^-] e^{-isJ} \varphi\, ds \\
&= \chi_0^- \varphi - i\, \lim_{\veps\downarrow 0} \int_{-\infty}^0 e^{\veps s} e^{isJ} [J,\chi_0^-]
e^{-isJ} \varphi\, ds
\end{align*}
Since the limit exists, we can replace $\veps$ by
$1/k$ and do the $s$ integral
\[
\begin{split}
(P_\ell^+  \varphi)_n  = \chi_0^- (n) \varphi_n  - &\lim_{k\to\infty}\, \sum_{m,n=-\infty}^\infty
\int G_{nm} \biggl( \lambda+\f{i}{k}\biggr)^{-1}  \\
& \qquad [J,\chi_0^+]_{m\ell} g(\lambda) u_\ell^+ (\lambda+i0)
f_+(\lambda)\, d\lambda
\end{split}
\]

But $[J,\chi_0^-]$ is rank two. In fact, $[J,\chi_0^+]_{m\ell}\neq
0$ only for $(m,\ell) =(0,1)$ or $(1,0)$, so the sum is finite, and
by the uniform convergence of $G_{nm} (\lambda+ \f{i}{k})$ for
$\lambda\in \supp(g)$ and $u_\ell^+ \in L^2(\bbR, f_+\, d\lambda)$,
we see that we can take the limit inside the integral. The result is
\begin{equation} \lb{2.29}
\begin{split}
(P_\ell^+ \varphi)_n & = \chi_0^- (n) \varphi_n - \int a_0 g(\lambda) \\
& \qquad [G_{n1}(\lambda+i0) u_0^+ (\lambda+i0) - G_{n0}
(\lambda+i0) u_1^+ (\lambda+i0)] f_+(\lambda)\, d\lambda
\end{split}
\end{equation}
If $n>0$, using \eqref{1.20},
\begin{align}
a_0 & [G_{n1} (\lambda+i0)  u_0^+ (\lambda+i0) - G_{n0} (\lambda+i0) u_1^+ (\lambda+i0)] \notag \\
& = \f{a_0 [u_0^+ (\lambda+i0) u_1^- (\lambda+i0) - u_1^+ (\lambda+i0) u_0^- (\lambda+i0)]}{W(\lambda)}\, u_n^+ (\lambda+i0) \lb{2.30} \\
&= u_n^+ (\lambda+i0) \notag
\end{align}
so \eqref{2.29} says
\begin{equation} \lb{2.31}
(P_\ell^+ \varphi)_n =\varphi_n
\end{equation}

If $n\leq 0$, the $u_{0,1}^+$ in \eqref{2.30} becomes $u_{0,1}^-$ and $u_n^+$ becomes $u_n^-$,
so the factor in $[\quad]$ is zero, and again  \eqref{2.31} holds.
\end{proof}

\begin{remark} $\chi_0^-$ can be replaced by any $\chi_\ell^-$. So in the analog of \eqref{2.29}
(where $G_{n1},G_{n0}$ become $G_{n\ell+1} G_{n\ell}$), one can even take $\ell$ to be $n$-dependent.
Using this, one can use either the argument we used for $n>0$ (by picking $\ell <n$) or for $n\leq 0$
(by picking $\ell \geq n$) rather than needing both calculations!
\end{remark}

The above implies $P_\ell^+ P_\fre =P_r^- P_\fre$ if and only if for
a.e.\ $\lambda\in\fre$, $u_n^+ = \ol{u_n^-}$, which holds if and
only if, by \eqref{1.18}/\eqref{1.19}, \eqref{1.10} holds for a.e.\
$\lambda\in\fre$. Thus, one has Theorem~\ref{T1.3}.

The following proves Theorem~\ref{T1.5}, and thereby completes the proofs of the results stated in
Section~\ref{s1}.

\begin{theorem}\lb{T2.5} For a.e.\ $\lambda\in\Sigma_\ac^{(2)}$, we can write
\begin{equation} \lb{2.32}
u_n^+ (\lambda+i0) = \alpha(\lambda)\, \ol{u_n^+ (\lambda+i0)} +
\beta(\lambda)\, \ol{u_n^- (\lambda+i0)}
\end{equation}
and the function $R$ of \eqref{1.13} is given by
\begin{equation} \lb{2.33}
R(\lambda) = \abs{\alpha(\lambda)}^2
\end{equation}
Moreover, $R(\lambda)$ is given by \eqref{1.14}.
\end{theorem}

\begin{proof} For a.e.\ $\lambda\in\Sigma_\ac^{(2)}$, $\Ima u_n^+ (\lambda)<0$, $\Ima u_n^- (\lambda)>0$, so $u^\pm(\lambda)$
are linearly independent solutions of $Ju=\lambda u$. It follows
that \eqref{2.32} holds. If
\begin{equation} \lb{2.34}
\varphi = \int g(\lambda) u_n^+ (\lambda+i0) f_+(\lambda)\,
d\lambda\in\ran (P_\ell^+)
\end{equation}
then \eqref{2.32} implies that
\begin{equation} \lb{2.35}
(P_\ell^- \varphi)_n =\int g(\lambda) \alpha(\lambda)\, \ol{u_n^+
(\lambda+i0)}\, f_+(\lambda)\, d\lambda
\end{equation}
from which
\begin{equation} \lb{2.36}
\norm{P_\ell^- \varphi}^2 = \int \abs{\alpha(\lambda) g(\lambda)}^2
f_+(\lambda)\, d\lambda
\end{equation}
This implies \eqref{2.33}.

If
\begin{equation} \lb{2.37}
W(f,g) = a_0 (g_1 f_0  - f_1 g_0)
\end{equation}
then \eqref{2.32} implies
\begin{equation} \lb{2.38}
\alpha(\lambda) = \f{W(u_\bddot^+ (\lambda+i0), \ol{u_\bddot^-
(\lambda+i0)})} {W(\ol{u_\bddot^+ (\lambda+i0)}, \ol{u_\bddot^-
(\lambda+i0)})}
\end{equation}

Since
\begin{equation} \lb{2.39}
u_0^\pm =1 \qquad u_1^+ =-a_0 m_0^+ \qquad u_1^- = -(a_0 m_0^-)^{-1}
\end{equation}
\eqref{2.38} implies \eqref{1.19}.
\end{proof}

\section{The Schr\"odinger Case} \lb{s3}

In this section, we consider a Schr\"odinger operator on $\bbR$,
\begin{equation} \lb{3.1x}
H = -\f{d^2}{dx^2} + V(x)
\end{equation}
where $V$ is in $L^1_\loc$ and limit point at both $+\infty$ and $-\infty$, so $H$ is the usual
selfadjoint operator (see, e.g., \cite[Appendix~A]{S249}). Because it is limit point, there are,
for any $z\in\bbC_+$, unique solutions $u_\pm(x,z)$ obeying
\begin{gather}
 -u''+Vu = zu \lb{3.1} \\
u_\pm(0,z)=1 \lb{3.2} \\
u_\pm\in L^2 (0,\pm\infty) \lb{3.3}
\end{gather}
For Lebesgue a.e. $\lambda\in\bbR$,
\begin{equation} \lb{3.4}
\lim_{\veps\downarrow 0}\, u_\pm (x,\lambda+i\veps) \equiv u_\pm
(x,\lambda +i0)
\end{equation}
exists for all $x\in\bbR$. Moreover, $\Sigma_\ac^{(2)}$, the a.c.\ spectrum of multiplicity $2$,
is determined by
\begin{equation} \lb{3.5}
\Ima (\mp u'_\pm (0, \lambda+i0)) >0
\end{equation}
(it is always $\geq 0$) for a.e.\ $x\in\Sigma_\ac^{(2)}$, that is, positivity for both $u_+$ and $u_-$.

The Weyl $m$-functions (see \cite[Appendix~A]{S249}) are defined by
\begin{equation} \lb{3.6}
m^\pm (x,\lambda+i0) = \mp \biggl[ \f{u'(x, \lambda+i0)}{u(x,
\lambda+i0)}\biggr]
\end{equation}
and for $\lambda\in\bbC_+$ if $\lambda+i0$ is replaced by $\lambda$.
We define $m(\lambda)\equiv m(x=0,\lambda)$. The Green's function is
given by (for $x\leq y$)
\begin{equation} \lb{3.7}
G(x,y;\lambda) = \f{u_-(x,\lambda) u_+(x,\lambda)}{W(\lambda)}
\end{equation}
where
\begin{equation} \lb{3.8}
W(\lambda) = u_-(x,\lambda) u'_+(x,\lambda) - u'_- (x,\lambda) u_+
(x,\lambda)
\end{equation}
is $x$-independent so that
\begin{equation} \lb{3.9}
W(\lambda) = -(m^+(\lambda) + m^- (\lambda))
\end{equation}
and
\begin{equation} \lb{3.10}
G(x,x;\lambda) = -(m^+ (x,\lambda) + m^- (x,\lambda))^{-1}
\end{equation}

$H$ is called {\it spectrally reflectionless\/} on
$\fre\subset\Sigma_\ac^{(2)}$ if and only if for a.e.\
$\lambda\in\fre$ and all $x$,
\begin{equation} \lb{3.11}
m^+ (x,\lambda+i0) = - \ol{m^- (x, \lambda+i0)}
\end{equation}

As proven in Davies--Simon \cite{Da-Si}, if $\chi_y^\pm$ is the characteristic function of $[y,\pm\infty)$, then
\begin{equation} \lb{3.12}
P_\ell^\pm = \slim_{t\to\mp\infty}\, e^{itH} \chi_y^- e^{-itH} P_\ac
\end{equation}
exists and is $y$-independent. Indeed, $\chi_y^-$ can be replaced by any continuous function, $j$,
which goes to $1$ at $-\infty$ and $0$ at $+\infty$. If $\chi_y^-$ is replaced by $\chi_y^+$,
we get $P_r^\pm$. If $\calH_{\ell,r}^\pm$ is $\ran (P_{\ell,r}^\pm)$, then \eqref{1.2} and
\eqref{1.3} hold. If \eqref{1.7a} and \eqref{1.4} hold, we say $H$ is dynamically reflectionless
on $\calH$.

Following \cite{Da-Si}, the dynamic reflection probability is given by \eqref{1.12}/\eqref{1.13} with
$J$ replaced by $H$. The spectral reflection probability (see, e.g., Gesztesy--Nowell--P\"otz \cite{GNP}
or Gesztesy--Simon \cite{S260}) is given on $\Sigma_\ac^{(2)}$ by
\begin{equation} \lb{3.13}
\biggl| \f{m^+ (\lambda+i0) + \ol{m^-(\lambda+i0)}}{m^+ (\lambda+i0)
+ m^- (\lambda+i0)}\biggr|^2
\end{equation}
Our main theorems in this case are:

\begin{theorem}\lb{T3.1} $H$ is dynamically reflectionless on $\fre\in\Sigma_\ac^{(2)}$ if and only
if it is spectrally reflectionless.
\end{theorem}

\begin{theorem} \lb{T3.2} $R(\lambda)$ is given by \eqref{3.13}.
\end{theorem}

The proofs closely follow those of Section~\ref{s2}, so we settle for a series of remarks explaining
the differences:

1. \ $S$ is now defined as
\begin{equation} \lb{3.14}
S(x,y; \lambda) = \pi^{-1} \Ima G(x,y; \lambda+i0)
\end{equation}
and there is still a Stone formula like \eqref{2.2x}. One defines
\begin{equation} \lb{3.15}
f_\pm (\lambda) = \f{\Ima m^\pm (\lambda+i0)}{\pi \abs{m^+ (\lambda+i0)
+ m^- (\lambda+i0)}^2}
\end{equation}
One proves
\begin{equation} \lb{3.16}
\begin{split}
S(x,y;\lambda) &= \ol{u_+(x,\lambda+i0)}\, u_+ (y,\lambda+i0) f_+(\lambda) \\
& \qquad  + \ol{u_-(x,\lambda+i0)}\, u_- (y, \lambda+i0) f_-
(\lambda)
\end{split}
\end{equation}
The proof is the same as that of Theorem~\ref{T2.1}, except sums over $k$ become integrals over $w\in\bbR$.

\smallskip
2. \ Once one has \eqref{3.16}, one can develop eigenfunction expansions analogously to
Theorem~\ref{T2.3}. The one difference is that since $\delta(x)$ is not in $L^2$, we do not
have the analog of \eqref{2.15a}. However,
\begin{equation} \lb{3.17}
\Ima G(x,x;\lambda=i) = \int \f{\Ima G(x,x; \lambda+i0)}{\lambda^2 +
1}\, d\lambda
\end{equation}
which implies that
\begin{equation} \lb{3.18x}
\int_{\Sigma_\ac^{(2)}} \f{\abs{u^\pm (x, \lambda+i0)}^2}{\lambda^2
+1}\, f_\pm (\lambda)\, d\lambda <\infty
\end{equation}
and that suffices to define an inverse transform on $L^2
(\Sigma_\ac^{(2)}, d\lambda)$ functions of compact support.

\smallskip
3. \ As a preliminary to the next step, we note that if $\eta$ is a function of compact support
with a continuous derivative and $q$ is $C^\infty$, then by an integration by parts,
\begin{equation} \lb{3.18}
\int \eta(x) \biggl[ q(x) \f{d}{dx}\, \eta(x) + \f{d}{dx}\,( q \eta)(x)\biggr] \, dx =0
\end{equation}

\smallskip
4. \ In computing $(P_\ell^+\varphi)(x_0)$ for $x_0 <0$, we can compute $\lim_{t\to\infty} (e^{itH}
je^{-itH} \varphi)$ with a $C^\infty j$ which is $1$ if $x<0$ and $0$ if $x>1$. Thus, in following
the calculation in the proof of Theorem~\ref{T2.4}, we start with
\begin{equation} \lb{3.19}
(P_\ell^- \varphi) (x_0) =\varphi(x_0) - i\, \lim_{\veps\downarrow 0} \int_{-\infty}^0 e^{\veps s} (e^{isH}
[H,j] \varphi)(x_0)\, ds
\end{equation}
Since $[H,j]$ involves $j'$ and $j''$, we can instead write $F[H,j]F$ where $F$ is multiplication by a
$C^\infty$ function supported in $(x_0,2)$ which is $1$ on $[0,1]$. When we put in the eigenfunction
expansion, we get
\begin{equation} \lb{3.20}
\int u^- (x_0, \lambda+i0) f_+ (\lambda) g(\lambda) h(\lambda)\,
d\lambda
\end{equation}
where $h$ has the form of the left side of \eqref{3.18} with
\begin{equation} \lb{3.21}
\eta(x) = F(x) u^+ (x,\lambda+i0) \qquad q(x) = -j'(x)
\end{equation}
yielding $(P_\ell^- \varphi)(x_0) = \varphi(x_0)$ for $x_0 <0$. By shifting $j$ to the right, we get this
for all $x_0$ (as in the remark following the proof of Theorem~\ref{T2.4}).

\section{The CMV Case} \lb{s4}

The basic objects in this section are two-sided CMV matrices, $\calC$, depending on a sequence
$\{\alpha_n\}_{n=-\infty}^\infty$ of Verblunsky coefficients. One-sided CMV matrices appeared first in
the numeric matrix literature \cite{BDMMZ,Bern2,PrS00} and were rediscovered by the OPUC community
\cite{CMV}. Two-sided CMV matrices were defined first in \cite{OPUC1}, although related objects appeared
earlier in \cite{BHJ,GTep}. For further study, we mention \cite{S293,GZjlms,GZjat,BRZ}.

$\calC$ is defined as follows. Given $\alpha\in\bbD$, we let $\rho =(1-\abs{\alpha}^2)^{1/2}$ and we let
$\Theta(\alpha)$ be the $2\times 2$ matrix,
\[
\Theta(\alpha) = \begin{pmatrix}
-\alpha & \rho \\
\rho & \overline \alpha
\end{pmatrix}
\]
and let $\Theta_j$ be $\Theta$ acting on $\delta_{j-1}, \delta_{j}$ in $\ell^2(\bbZ)$. Then
\begin{equation} \lb{4.1}
\calC=\calL\calM
\end{equation}
where
\begin{equation} \lb{4.2}
\calL = \bigoplus_{n=-\infty}^\infty \Theta_{2n} (\alpha_{2n}) \qquad
\calM = \bigoplus_{n=-\infty}^\infty \Theta_{2n+1} (\alpha_{2n+1})
\end{equation}

First, one can develop a unitary analog of the Davies--Simon theory \cite{Da-Si}. It is not hard to show that
the Pearson theorem on two-space scattering (see, e.g., \cite[Thm.~XI.7]{RS3}) extends to the unitary case.  That is, if $U$ and $V$ are unitary,  $J$ is bounded, and $UJ-JV$ is trace class, then
\begin{equation} \lb{4.3}
\slim_{t\to\mp\infty} U^{-n} JV^n P_\ac(V)
\end{equation}
exists. Thus, if $\chi_n^\pm$ are defined as in Section~\ref{s1}, one defines
\begin{align}
P_\ell^\pm &= \slim_{n\to\mp\infty} \calC^{-n} \chi_0^- \calC^n P_\ac(\calC) \notag \\
P_r^\pm &=\slim_{n\to\pm\infty} \calC^{-n} \chi_0^+ \calC^n P_\ac (\calC) \lb{4.4}
\end{align}
As in Section~\ref{s1}, we define
\begin{equation} \lb{4.5}
\calH_{\ell,r}^\pm = \ran (P_{\ell,r}^\pm)
\end{equation}
and we say $\calC$ is {\it dynamically reflectionless\/} on $\fre$ if \eqref{1.7a} and \eqref{1.4} hold.

If $\alpha_{n-1}$ is replaced by $1$, the CMV matrix breaks into a direct sum of two CMV matrices, $\calC_n^+$
on $\ell^2 (\{n,n+1, \dots\})$ and $\calC_{n-1}^-$ on $\ell^2 (\{n-2, n-3, \dots\})$.  $F_+ (z,n)$ is defined
for $z \not\in \partial\bbD$ by setting
\begin{equation} \lb{4.6}
F_+(z,n) = \biggl\langle \delta_n, \biggl( \f{\calC_n^+ +z}{\calC_n^+ -z}\biggr) \delta_n\biggr\rangle
\end{equation}
and $F_- (z,n-1)$ by
\begin{equation} \lb{4.7}
F_-(z,n-1) = \biggl\langle \delta_{n-1}, \biggl( \f{\calC_{n-1}^- + z}{\calC_{n-1}^- -z}\biggr) \delta_{n-1}\biggr\rangle
\end{equation}
It is known (see, e.g., \cite{OPUC1}) that  when restricted to $z \in \mathbb D$, $F_+ (z,n)$ is the Carath\'{e}odory
function whose Verblunsky coefficients are $\{\alpha_n, \alpha_{n+1}, \dots\}$ and $F_- (z,n-1)$ has Verblunsky coefficients
$\{-\bar\alpha_{n-2}, - \bar\alpha_{n-3}, \dots\}$.  We will let $F_\pm(z) = F_\pm(z,n=0)$.

As Carath\'{e}odory functions, $F_\pm (z,n)$ have a.e.\ boundary values on $\partial\bbD$ which we denote by
$F_\pm (e^{i\theta},n) = \lim_{r \uparrow 1} F_\pm(r e^{i\theta},n)$. $\calC$ is called {\it spectrally reflectionless\/}
on $\fre\subset\partial\bbD$ if and only if for a.e.\ $e^{i\theta}\in\fre$ and all $n\in\bbZ$,
\begin{equation} \lb{4.8}
F_+(e^{i\theta} , n) = \overline{ F_- (e^{i\theta},n)}
\end{equation}

There is an equivalent definition using Schur functions (see, e.g.,
\cite{BRZ}). The equivalence is an easy computation using the
relations between the Carath\'{e}odory and Schur functions (see,
e.g., \cite{GZjlms}). It is known \cite{GZjlms} that \eqref{4.8} for
one $n$ implies it for all $n$. It is also known \cite{BRZ} that
while \eqref{4.8} implies $\jap{\delta_n,
(\calC+z)/(\calC-z)\delta_n}$ has purely real boundary values a.e.\
on $\fre$, the converse can be false.

The dynamic reflection probability $R(e^{i\theta})$ is given by \eqref{1.12}/\eqref{1.13} with
$J$ replaced by $\mathcal C$. The spectral reflection probability is given on $\Sigma_\ac^{(2)}$ by
\begin{equation}\label{cmv ref prob}
\biggl| \frac{F_+(e^{i\theta}) - \overline{F_-(e^{i\theta})} }{ F_+(e^{i\theta}) + F_-(e^{i\theta})}  \biggr|^2
\end{equation}

Our main theorems in this case are:
\begin{theorem} \lb{T4.1} $\calC$ is dynamically reflectionless on $\fre$ if and only if it is spectrally
reflectionless on $\fre$.
\end{theorem}

\begin{theorem}\lb{T4.2} $R(e^{i\theta})$ is given by \eqref{cmv ref prob}.
\end{theorem}

The proofs closely follow those of Section \ref{s2}, so we again settle for a series of remarks:

1. \ The analysis requires us to simultaneously study solutions of $\mathcal C$ and $\mathcal C^T$.  To do so, let
$$
\mathcal E = \begin{pmatrix} \mathcal C & 0 \\ 0 & \mathcal C^T \end{pmatrix}
$$
acting on two sequences labeled by all of $\bbZ$. Following
Gesztesy--Zinchenko \cite{GZjat}, let
$$
\begin{pmatrix} p(z,n) \\ r(z,n) \end{pmatrix} \quad\text{ and }\quad \begin{pmatrix} q(z,n) \\ s(z,n)\end{pmatrix}
$$
be the two (Laurent polynomial) solutions to the equation
\begin{equation}\label{E equation}
\mathcal E \begin{pmatrix} u \\ v \end{pmatrix} = z \begin{pmatrix} u \\ v \end{pmatrix}
\end{equation}
satisfying the initial conditions
$$
\begin{pmatrix} p(z,0) \\ r(z,0) \end{pmatrix} = \begin{pmatrix} 1 \\ 1 \end{pmatrix}
\quad\text{ and }\quad \begin{pmatrix} q(z,0) \\ s(z,0) \end{pmatrix} = \begin{pmatrix} -1 \\ 1 \end{pmatrix}
$$
That is, for one solution, the components of $u$ are $p$ and of $v$
are $r$, and this solution is uniquely determined by the initial
conditions given (see \eqref{transfer equation}).  Similarly, the
components for the second solution are given by $q$ and $s$. Finally, we let
$$
\begin{pmatrix} u_\pm (z,n) \\ v_\pm (z,n) \end{pmatrix} = \begin{pmatrix} q(z,n) \\ s(z,n) \end{pmatrix}
\pm F_\pm (z) \begin{pmatrix} p(z,n) \\ r(z,n) \end{pmatrix}
$$
be the unique solutions that are $\ell^2$ at $\pm \infty$, normalized by
\begin{equation}\label{normalization cmv}
\begin{pmatrix} u_\pm (z,0) \\ v_\pm (z,0) \end{pmatrix} =
\begin{pmatrix}-1\pm F_\pm(z) \\ 1 \pm F_\pm(z) \end{pmatrix}
\end{equation}

We note that there are a number of relations between $u_\pm$ and $v_\pm$ that we will need (see \cite{GZjat}).
First, \eqref{E equation} is equivalent to
\begin{equation}\label{transfer equation}
\begin{pmatrix} u(z,n) \\ v(z,n) \end{pmatrix} = T(z,n) \begin{pmatrix} u(z,n-1) \\ v(z, n-1)\end{pmatrix}
\end{equation}
where
\begin{equation}\label{transfer matrix}
T(z,n) = \begin{cases} \dfrac{1}{\rho_n} \begin{pmatrix} \alpha_n & z \\ 1/z & \overline{\alpha_n}\end{pmatrix} ,
& n \text{ odd}\\
\dfrac{1}{\rho_n} \begin{pmatrix} \overline{\alpha_n} & 1 \\ 1 & \alpha_n\end{pmatrix} ,
& n \text{ even} \end{cases}
\end{equation}

Similarly, \eqref{E equation} implies
\begin{align}\label{theta equation}
\begin{split}
\begin{pmatrix} u(z,2n-1) \\ u(z,2n) \end{pmatrix} &= \Theta_{2n}(\alpha_{2n}) \begin{pmatrix} v(z,2n-1) \\ v(z,2n) \end{pmatrix}\\
\begin{pmatrix} u(z,2n-2) \\ u(z,2n-1) \end{pmatrix} &= \Theta_{2n-1}(\alpha_{2n-1}) \begin{pmatrix} v(z,2n-2) \\ v(z,2n-1) \end{pmatrix}
\end{split}
\end{align}

Finally, for all $n \in \mathbb Z$, we have
\begin{equation}\label{uv relation}
v_\pm (1/\bar z , n ) = -\overline{ u_\pm (z,n)}
\end{equation}
This is because
$$
\mathcal C u = z u \text{ holds } \quad \Leftrightarrow\quad \mathcal C^T \bar u = (1/\bar z) \bar u \text{  holds}
$$
and \eqref{4.6}/\eqref{4.7} imply $F_\pm(1/\bar z) = -\overline{ F_\pm (z)}$, and because the solutions to
\eqref{E equation} that are $\ell^2$  at $\pm \infty$ are unique up to normalization.

\smallskip

2. \ Using the solutions $u_\pm(z,n)$ and $v_\pm(z,n)$ we can write the analog of \eqref{1.20} (see \cite{GZjat}):
\begin{equation}\label{uv expansion}
(\mathcal C - z)^{-1}_{nm} = \frac{-1}{zW(z)} \begin{cases} u_-(z,n)v_+(z,m), \quad n<m \text{ or } n = m=2k+1\\
v_-(z,m) u_+(z,n), \quad m<n \text{ or } n = m=2k \end{cases}
\end{equation}
where
\begin{equation}\label{Wronskian}
W(z) = u_+(z,n)v_-(z,n) - v_+(z,n)u_-(z,n)
\end{equation}
is independent of $n \in \mathbb Z$.

\smallskip

3. \ Next we find the analog of $[J , \chi_0^+]$.  Due to the structure of \eqref{4.1}, the results are different
depending on whether $n$ is even or odd. For $n$ even:
\begin{equation}\label{commutator}
[\mathcal C , \chi_n^+] = -\rho_{n} \bigl(  \rho_{n-1} |\delta_{n} \rangle \langle \delta_{n-2} |   + \overline{\alpha_{n-1}} | \delta_{n} \rangle \langle \delta_{n-1}|    +  \alpha_{n+1} | \delta_{n-1} \rangle \langle \delta_{n}|      - \rho_{n+1} | \delta_{n-1} \rangle \langle \delta_{n+1}|\bigr)
\end{equation}
while if $n$ is odd we get the same thing but transposed and with a minus sign:
\begin{equation}\label{commutator 2}
[\mathcal C , \chi_n^+] = \rho_{n} \bigl(  \rho_{n-1} |\delta_{n-2} \rangle \langle \delta_{n} |   + \overline{\alpha_{n-1}} | \delta_{n-1} \rangle \langle \delta_{n}|    +  \alpha_{n+1} | \delta_{n} \rangle \langle \delta_{n-1}|      - \rho_{n+1} | \delta_{n+1} \rangle \langle \delta_{n-1}|\bigr)
\end{equation}

\smallskip

4. \ $S$ is defined (using a.e.\ boundary values) as
\begin{equation}\label{cmv S}
S(n,m; e^{i\theta}) = \f{1}{2\pi}\, \lim_{r \uparrow 1} ( (\mathcal C + re^{i\theta})(\mathcal C - re^{i\theta})^{-1} - (\mathcal C + r^{-1}e^{i\theta})(\mathcal C - r^{-1}e^{i\theta})^{-1})_{n m}
\end{equation}
and there is a Stone formula like \eqref{2.2x}. Proceeding as in Section \ref{s2} and using \eqref{uv relation} and \eqref{uv expansion}, one can deduce the analog of \eqref{2.7}:
\begin{align}\label{cmv s}
S(n,m;e^{i\theta}) =  u_+(e^{i\theta},n) \overline{ u_+(e^{i\theta},m)} f_+(e^{i\theta}) + u_-(e^{i\theta},n) \overline{ u_-(e^{i\theta},m)} f_-(e^{i\theta})
\end{align}
where $u_\pm(e^{i\theta},n) = \lim_{r \uparrow 1} u_\pm(r e^{i\theta},n)$ and
$$
\pi f_\pm(e^{i\theta}) = \lim_{r \uparrow 1} \frac{1}{r e^{i\theta}|W(r e^{i\theta})|^2}  \langle
u_\mp(r^{-1} e^{i\theta}) , [\mathcal C , \chi_k^\mp ] u_\mp(r^{-1} e^{i\theta})   \rangle
$$

As before, this is independent of $k$, and choosing $k=0$ one may use \eqref{normalization cmv}, \eqref{transfer equation}, and \eqref{commutator}/\eqref{commutator 2} to find the analog of \eqref{2.2}:
\begin{equation}\label{f cmv}
f_\pm(e^{i\theta}) = \frac{4 \Real F_\mp(e^{i\theta})}{\pi|W(e^{i\theta})|^2}
\end{equation}

Once one has \eqref{cmv s}, one may develop eigenfunction expansions exactly as in Theorem \ref{T2.3}.

\smallskip

5. \ To prove Theorem \ref{T4.1}, we first define $P^\pm$ and $\overline{ P^\pm}$ as in Section \ref{s2} but using
$\lim_{r \uparrow 1} u_\pm( r e^{i\theta},n)$ and $\lim_{r \uparrow 1} u_\pm( r^{-1} e^{i\theta},n)$ respectively.
As before, we consider $P_\ell^+ = \slim_{n \rightarrow -\infty} \mathcal C^{-n} \chi_0^- \mathcal C^n P_\ac(\mathcal C)$.  Because
$$
\mathcal C^{-n} \chi_0^- \mathcal C^n - \mathcal C^{-(n-1)} \chi_0^- \mathcal C^{n-1} = \mathcal C^{-n}[\chi_0^- , \mathcal C ] C^{n-1}
$$
and the strong limit defining $P_\ell^+$ exists, we see
$$
P_\ell^+ = \chi_0^- + \slim_{n \rightarrow -\infty} \sum_{k=1}^n \mathcal C^{-k} [\chi_0^- , \mathcal C] \mathcal C^{k-1}
$$
Choosing a dense set of $\varphi \in \ran (P^+)$ as before, and using the eigenfunction expansion and an
abelian theorem, we find
\begin{align*}
(P_\ell^+ \varphi)_m &= (\chi_0^- \varphi)_m + \Bigl(\lim_{n \rightarrow -\infty} \sum_{k=1}^n \mathcal C^{-k} [\chi_0^- , \mathcal C] \mathcal C^{k-1} \varphi \Bigr)_m\\
&=(\chi_0^- \varphi)_m + \lim_{r \uparrow 1} \lim_{n \rightarrow -\infty}  \int \sum_{k=1}^n \mathcal C^{-k} [\chi_0^- , \mathcal C] (r e^{i\theta})^{k-1}  u_+(e^{i\theta},m) g(e^{i\theta}) f_+(e^{i\theta}) \frac{d\theta}{2\pi}\\
&=(\chi_0^-\varphi)_m + \int (\mathcal C - e^{i\theta})^{-1} [\chi_0^- , \mathcal C] u_+(e^{i\theta},m) g(e^{i\theta}) f_+(e^{i\theta}) \frac{d\theta}{2\pi}
\end{align*}

The proof of Theorem \ref{T4.1} then proceeds exactly as the proof of Theorem \ref{T1.3}, but now using \eqref{normalization cmv}--\eqref{f cmv}.  The proof of Theorem \ref{T4.2} follows that of Theorem \ref{T2.5} but with $\lim_{r \uparrow 1}u_\pm(re^{i\theta},n)$ and $\lim_{r \uparrow 1} u_\pm(r^{-1} e^{i\theta},n)$ replacing $u_\pm(x+i0,n)$ and $\overline{ u_\pm (x+i0,n)}$.

\bigskip

\end{document}